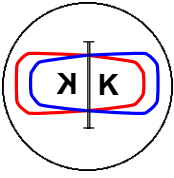

**DAΦNE TECHNICAL NOTE**

INFN - LNF, Accelerator Division

# Strong RF Focusing for Luminosity Increase

*A. Gallo, P. Raimondi and M. Zobov*

## 1. Introduction

The minimum value of the vertical beta-function $\beta_y$ at the IP in a collider is set by the hourglass effect [1] and it is almost equal to the bunch length $\sigma_z$. Reduction of the bunch length is an obvious approach to increase the luminosity. By scaling the horizontal and vertical beta functions $\beta_x$ and $\beta_y$ at the IP as the bunch length $\sigma_z$, the linear tune shift parameters $\xi_{x,y}$ remain unchanged while the luminosity scales as $1/\sigma_z$ [2]:

$$L \propto \frac{1}{\sigma_x \sigma_y} \propto \frac{1}{\sqrt{\beta_x \beta_y}} \propto \frac{1}{\sigma_z} \qquad (1)$$

A natural way to decrease the bunch length is to decrease the storage ring momentum compaction and/or to increase the RF voltage. However, in such a way we cannot obtain very short bunches since the short-range wakefields prevent this because of the potential well distortion and microwave instability.

In this paper we consider an alternative strategy to get short bunches at the IP. In particular, we propose to use strong RF focusing [3] (with high RF voltage and high momentum compaction) to obtain very short bunches at the IP with progressive bunch elongation toward the RF cavity.
With respect to the case of short bunches with constant length all along the ring, the situation seems more comfortable since the average charge density driving the Touschek scattering is smaller. Besides, this allows placing the most important impedance generating devices near the RF cavity where the bunch is longest thus minimizing the effect of the wakefields.

## 2. Strong RF Focusing

In order to compress the bunch at the IP in a collider a strong RF focusing can be applied. For this purpose high values of the momentum compaction factor $\alpha_c$ and extremely high values of the RF gradient are required. It is estimated that, for a Φ-factory collider, an RF voltage $V_{RF}$ of the order of 10 MV is necessary provided that the $\alpha_c$ value is of the order of 0.2.

Under these conditions the synchrotron tune $v_s$ grows to values larger than 0.1 and the commonly used "smooth approximation" in the analysis of the longitudinal dynamics is no longer valid. Instead, the longitudinal dynamics is much more similar to the transverse one, and can be analyzed on the base of the simple linear model reported in Fig. 1 and of its transfer matrices. In this model the cavity behaves like a thin focusing lens in the longitudinal phase space, while the rest of the machine is a drift space, where the "drifting" variable is the $R_{56}(s)$. In Fig. 1 $\lambda_{RF} = c/f_{RF}$ is the RF wavelength, $E/e$ is the particle energy in voltage units, while $L$ is the total ring length.

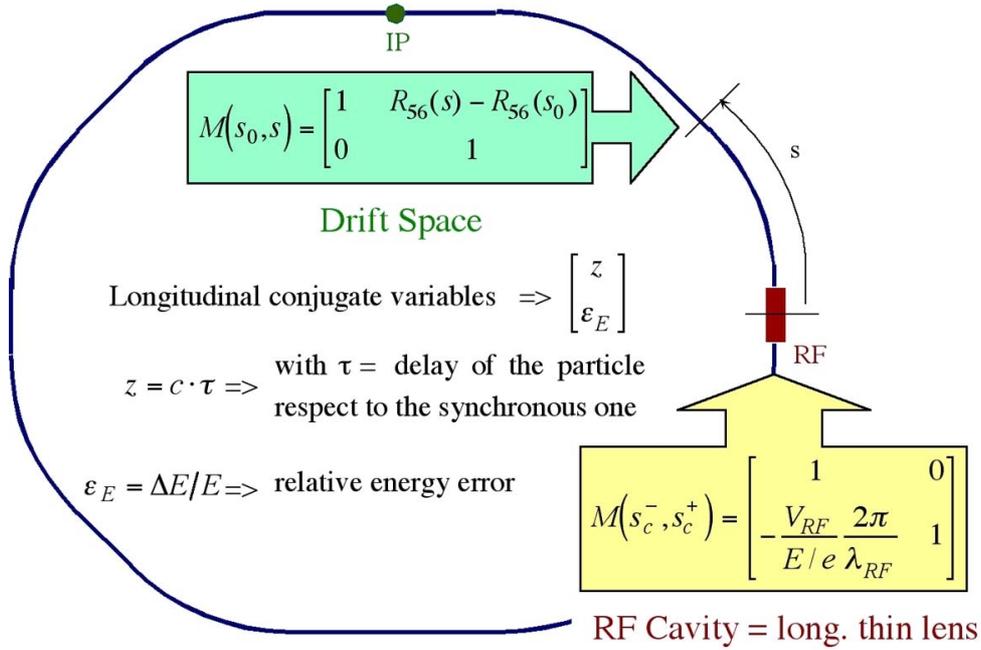

Figure 1: Linear model for the analysis of the longitudinal dynamics

The $R_{56}(s)$ parameter relates the path length to the normalized energy deviation of a particle, and it is given by:

$$R_{56}(s) = \int_0^s \frac{\eta(\tilde{s})}{\rho(\tilde{s})} \, d\tilde{s} \tag{2}$$

where $\rho(s)$ is the local bending radius and $\eta(s)$ is the ring dispersion function.

Taking the cavity position as the reference point $s = 0$, the one-turn transfer matrix $M(s, s+L)$ of this system starting from the generic azimuth $s$ is given by:

$$M(s,s+L) = \begin{bmatrix} 1 - 2\pi \dfrac{R_{56}(s)}{\lambda_{RF}} \dfrac{V_{RF}}{E/e} & \alpha_c L \left(1 - 2\pi \dfrac{R_{56}(s)}{\lambda_{RF}} \left(1 - \dfrac{R_{56}(s)}{\alpha_c L}\right) \dfrac{V_{RF}}{E/e}\right) \\ -\dfrac{V_{RF}}{E/e} \dfrac{2\pi}{\lambda_{RF}} & 1 + 2\pi \dfrac{R_{56}(s)}{\lambda_{RF}} \left(1 - \dfrac{\alpha_c L}{R_{56}(s)}\right) \dfrac{V_{RF}}{E/e} \end{bmatrix} \tag{3}$$

The one turn synchrotron phase advance is given by:

$$\cos\mu = \frac{1}{2}Tr[M(s,s+L)] = 1 - \pi \frac{\alpha_c L}{\lambda_{RF}} \frac{V_{RF}}{E/e} \quad (4)$$

leading to the following stability condition:

$$|\cos\mu| \leq 1 \implies \mu \leq \pi \implies \nu_s \leq 1/2 \implies V_{RF} \leq \frac{2}{\pi} \frac{\lambda_{RF}}{\alpha_c L} E/e = V_{RF_{Max}} \quad (5)$$

which shows that there is a constraint in the choice of the values of $V_{RF}$ and $\alpha_c$.

The one-turn transfer matrix can be put in canonical form:

$$M(s,s+L) = \cos\mu \cdot \hat{I} + \sin\mu \cdot \hat{J} = \cos\mu \cdot \begin{bmatrix} 1 & 0 \\ 0 & 1 \end{bmatrix} + \sin\mu \cdot \begin{bmatrix} \alpha_l & \beta_l \\ -\gamma_l & -\alpha_l \end{bmatrix} \quad (6)$$

and the longitudinal Twiss parameters are given by:

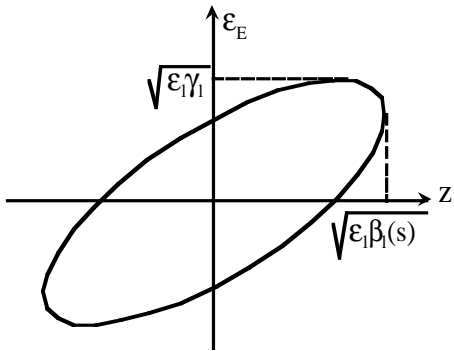

Longitudinal phase space ellipse

$$\alpha_l(s) = \frac{\pi}{\sin\mu} \frac{V_{RF}}{E/e} \frac{\alpha_c L - 2R_{56}(s)}{\lambda_{RF}}$$

$$\beta_l(s) = \frac{\alpha_c L}{\sin\mu}\left(1 - 2\pi \frac{R_{56}(s)}{\lambda_{RF}}\left(1 - \frac{R_{56}(s)}{\alpha_c L}\right)\frac{V_{RF}}{E/e}\right) \quad (7)$$

$$\gamma_l(s) = \frac{1}{\sin\mu} \frac{V_{RF}}{E/e} \frac{2\pi}{\lambda_{RF}}$$

Since $\gamma_l$ does not depend upon $s$, the vertical size of the ellipse (which represent the normalized energy spread $\sigma_E/E$ of the equilibrium distribution) does not vary along the ring. The longitudinal emittance $\varepsilon_l$ is related to the equilibrium energy spread according to:

$$\sigma_E/E = \sqrt{\varepsilon_l \gamma_l} \implies \varepsilon_l = (\sigma_E/E)^2 \frac{\sin\mu}{2\pi} \frac{E/e}{V_{RF}} \lambda_{RF} = (\sigma_E/E)^2 \frac{\sin\mu}{1-\cos\mu} \frac{\alpha_c L}{2} \quad (8)$$

On the contrary, since $\beta_l$ does depend upon $s$, the horizontal size of the ellipse (i.e. the bunch length $\sigma_z$) varies along the ring according to:

$$\sigma_z(s) = \sqrt{\varepsilon_l \beta_l(s)} = (\sigma_E/E)\sqrt{\frac{\alpha_c L}{2\pi} \frac{E/e}{V_{RF}} \lambda_{RF}\left(1 - 2\pi \frac{R_{56}(s)}{\lambda_{RF}}\left(1 - \frac{R_{56}(s)}{\alpha_c L}\right)\frac{V_{RF}}{E/e}\right)} =$$

$$= \sigma_z(0)\sqrt{1 - 2\pi \frac{R_{56}(s)}{\lambda_{RF}}\left(1 - \frac{R_{56}(s)}{\alpha_c L}\right)\frac{V_{RF}}{E/e}} \quad (9)$$

where $\sigma_z(0)$ is the bunch length at $s=0$ (i.e. at the cavity position). It may be noticed that $\sigma_z(0) = \sigma_{z_{max}}$ is the maximum value of the bunch length along the ring. On the other hand, the minimum value $\sigma_{z_{min}}$ corresponds to the $s_{min}$ position where $R_{56}(s_{min}) = \alpha_c L/2$.

If the position of the minimum corresponds to the IP one gets:

$$\sigma_z(IP) = \sigma_z(Cav) \sqrt{1 - \frac{\pi}{2} \frac{\alpha_c L}{\lambda_{RF}} \frac{V_{RF}}{E/e}} = \sigma_z(Cav) \sqrt{\frac{1+\cos\mu}{2}} \quad (10)$$

As $\mu$ approaches 180°, the ratio between the bunch lengths at the IP and at the RF goes to zero. This result is of great interest since it allows designing a ring where the bunch is short at the IP and progressively elongates moving toward the RF position.

### 3. Equilibrium Energy spread

In order to compute exactly the bunch size along the ring one needs to know the longitudinal emittance value (or, equivalently, the value of the equilibrium energy spread). These values can be worked out from a rigorous analysis of the longitudinal dynamics (abandoning the smooth approximation) or from a multi-particle tracking simulation including the distributed emission process along the machine. We follow an analytical approach based on the computation of the second momenta of the bunch equilibrium distributions using the eigenvectors of the longitudinal one-turn transfer matrix [4] that gives the following result:

$$\left(\frac{\sigma_E}{E}\right)^2 = \frac{1}{1+\cos\mu} \frac{55}{48\sqrt{3}} \frac{r_e \hbar}{m_e} \frac{\gamma^5 \tau_d}{L} \oint \left[1 - \frac{2\pi R_{56}(s)}{\lambda_{RF}}\left(1 - \frac{R_{56}(s)}{\alpha_c L}\right)\frac{V_{RF}}{E/e}\right] \frac{ds}{|\rho(s)|^3} \quad (11)$$

where $r_e$ and $m_e$ are the electron classical radius and rest mass, $\tau_d$ is the longitudinal damping time and $\gamma = E/(m_e c^2)$ is the relativistic factor. It may be noticed that the equilibrium energy spread $\frac{\sigma_E}{E}$ is diverging as $\mu$ tends to 180°, while at low tunes it tends to the value $\left.\frac{\sigma_E}{E}\right|_0$:

$$\left(\left.\frac{\sigma_E}{E}\right|_0\right)^2 = \frac{55}{96\sqrt{3}} \frac{r_e \hbar}{m_e} \frac{\gamma^5 \tau_d}{L} \oint \frac{ds}{|\rho(s)|^3} \quad (12)$$

which is the expression commonly reported in literature [5].

Expression (12) can be also conveniently rewritten in the following forms:

$$\left(\frac{\sigma_E}{E}\right)^2 = \frac{1}{1+\cos\mu} \frac{55}{48\sqrt{3}} \frac{r_e \hbar}{m_e} \frac{\gamma^5 \tau_d}{L} \oint \frac{\beta_l(s)}{\beta_l(0)|\rho(s)|^3} ds =$$

$$= \left(\left.\frac{\sigma_E}{E}\right|_0\right)^2 \frac{2}{1+\cos\mu} \frac{\oint \frac{\beta_l(s)}{\beta_l(0)|\rho(s)|^3} ds}{\oint \frac{ds}{|\rho(s)|^3}} \quad (13)$$

In the simplified assumption of constant bending radius $\rho$ and $R_{56}(s)$ linearly growing in the arcs, expression (12) reduces to:

$$\left(\frac{\sigma_E}{E}\right)^2 = \frac{2}{3}\left(\left.\frac{\sigma_E}{E}\right|_0\right)^2 \frac{2+\cos\mu}{1+\cos\mu} \quad (14)$$

Different results may be obtained if the ring has variable bending radii and/or the $R_{56}(s)$ function does not grow linearly in the arcs.

Under the assumptions leading to eq. (14), the longitudinal emittance $\varepsilon_l$ and the bunch lengths at the RF cavity and IP are given by:

$$\varepsilon_l = \frac{\alpha_c L}{3}\left(\left.\frac{\sigma_E}{E}\right|_0\right)^2 \frac{2+\cos\mu}{\sin\mu}$$

$$\sigma_z(Cav) = \frac{\alpha_c L}{\sin\mu}\left(\left.\frac{\sigma_E}{E}\right|_0\right)\sqrt{\frac{2+\cos\mu}{3}}; \quad \sigma_z(IP) = \alpha_c L\left(\left.\frac{\sigma_E}{E}\right|_0\right)\sqrt{\frac{2+\cos\mu}{6(1-\cos\mu)}}$$

(15)

The emittance and the bunch length at the RF cavity, as well as the energy spread, diverge as $\mu$ approaches 180°, while the bunch length at the IP remains finite

Figure 2 shows the longitudinal emittance and the equilibrium energy spread as a function of the phase advance $\mu$. The lines correspond to the analytical expressions (14)-(15), while dots represent the results of the multi-particle tracking simulations. The bunch length dependences on $\mu$ (both analytical and numerical) are reported in Fig. 3.

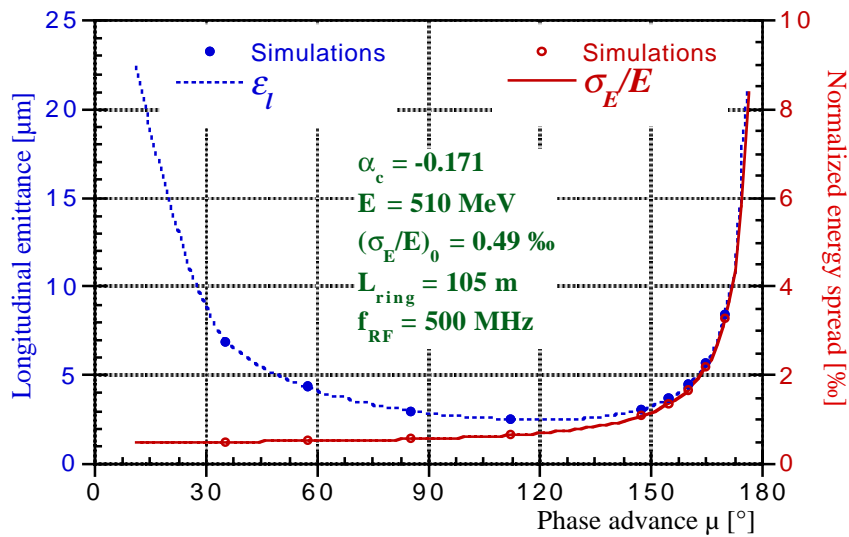

Figure 2: Longitudinal emittance and energy spread vs. phase advance

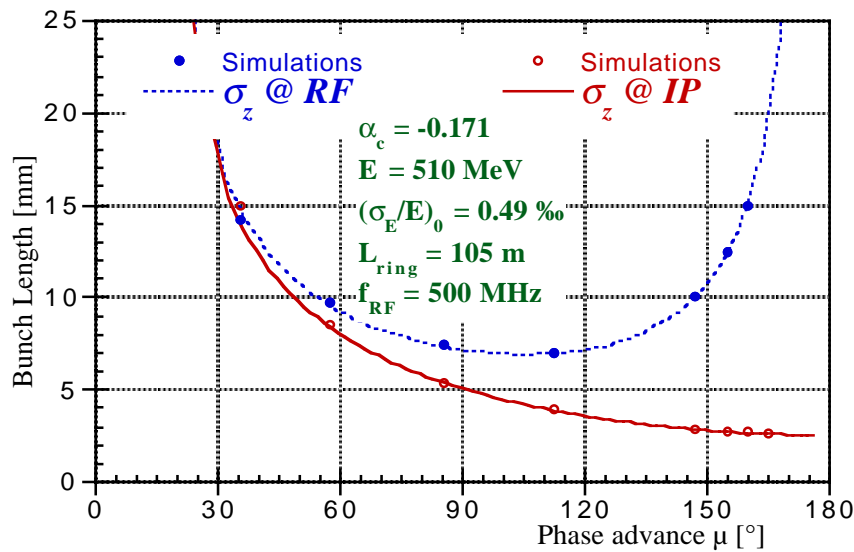

Figure 3: Bunch length @ RF and IP vs. phase advance

As it is seen, the longitudinal emittance exhibits a minimum at $\mu = 120°$. The analogy with the transverse case is quite evident [6]. Being the momentum compaction fixed, the various phase advances correspond to different values of the RF voltages. The voltage required to approach the limit phase advance value of 180° exceeds 10 MV. The use of superconducting cavities is mandatory in this case.

## 4. Conclusions

In this paper the motion of particles in a strong longitudinal focusing storage ring is described by means of the linear matrices formalism. Longitudinal optical functions are derived, showing that the bunch length varies along the ring and may be minimized at the IP. Analytical expressions for the longitudinal emittance and the energy spread of the bunch equilibrium distribution have been obtained and validated by comparison with results from multiparticle tracking simulations. It has been shown that the longitudinal emittance and the energy spread, as well as the bunch length at the RF cavity position diverge as the synchrotron phase advance approaches 180° per turn, while the bunch length at the IP tends to a minimum value which is finite.

Many aspects of beam physics need to be studied to establish whether or not a collider may efficiently work in the strong longitudinal focusing regime. The most relevant issues are bunch lengthening due to the wakefields, Touschek lifetime, dynamic aperture and beam-beam effect. Very preliminary multiparticle tracking simulations based on the DAFNE short range wake show that the short bunch length at the IP can be preserved up to relatively high bunch current (> 10 mA) provided that all the wake is concentrated near the RF cavity, the position where the bunch is longest.